\documentclass[twocolumn,secnumarabic,amssymb, nobibnotes, aps, prd]{revtex4-1}

\usepackage{graphicx}
\usepackage{graphics}
\usepackage{subfigure}

\begin{document}

\title {Ralph's equivalent circuit model, revised Deutsch's maximum entropy rule and discontinuous quantum evolutions in D-CTCs}

\author{Xiao Dong, Hanwu Chen, Ling Zhou}

\affiliation{Faculty of Computer Science and Engineering, Southeast University, Nanjing, China}

\email{Xiao.Dong@seu.edu.cn}


\begin{abstract}
We examine Ralph's equivalent circuit model of D-CTCs, which was proposed to derive Deutsch's maximum entropy rule of D-CTCs. By constructing counterexamples we show that the equivalent circuit model does not always reproduce the unique fixed state with the maximal von Neumann entropy. We speculate that the equivalent circuit model remains the correct description of D-CTCs and it can reproduce a revised maximum entropy rule of D-CTCs. We also suggest that the revised maximum entropy rule may eliminate the discontinuous quantum evolutions of D-CTCs.
\end{abstract}

\maketitle

\section{Introduction}
Closed timelike curves (CTCs) bring many new features in quantum information processing, such as state cloning, discrimination of non-orthogonal states, violation of Heisenberg's uncertainty relationship and information processing power beyond standard quantum mechanism\cite{PhysRevLett.111.190401}\cite{brun2012perfect}\cite{pienaar2013open}. There exist two popular but inequivalent models of CTCs, Deutch's D-CTCs and Lloyd's P-CTCs.  But both of them suffer from paradoxes due to either dynamical consistency problems or information paradoxes\cite{PhysRevA.82.062330}\cite{allen2014treating}.

This paper will address the uniqueness ambiguity problem of D-CTCs, which is the origin of the information paradoxes of D-CTCs.

 In the standard quantum circuit form of D-CTCs\cite{deutsch1991quantum}, we have both the qubits that can travel back in time through CTCs called \emph{chronology violating} (CV) qubits and those do not travel back called \emph{chronology respecting} (CR) qubits. The CV and CR systems interact with each other by an unitary quantum interaction $U$. Then Deutsch's consistent solution of the CV system is determined by
\begin{equation}\label{eq-1}
\tau_{CV}=D_{U}(\rho_{CR}^{in},\tau_{CV})=Tr_{CR}(U(\rho_{CR}^{in}\otimes\tau_{CV})U^{\dagger})
\end{equation}

and the output of the CR system is given by
\begin{equation}\label{eq-2}
\rho_{CR}^{out}=Tr_{CV}(U(\rho_{CR}^{in}\otimes\tau_{CV})U^{\dagger})
\end{equation}

Deutsch proved that at least one consistent solution of Eq. (\ref{eq-1}) exists for every $U$ and $\rho_{CR}^{in}$ in the form of a density operator, which is
\begin{equation}\label{eq_3}
\overline{D_{U}^{\infty}}(\rho_{CR}^{in},\tau_{CV}^{0})=\lim_{n\rightarrow\infty}\frac{1}{n+1}\sum_{k=0}^{n}D_{U}^{k}(\rho_{CR}^{in},\tau_{CV}^{0})
\end{equation}
where $D_{U}^{k}(\rho_{CR}^{in},\tau_{CV}^{0})$ is defined to be $k$ consecutive applications of the interaction $U$.

The consistent solution Eq. (\ref{eq_3}) may be dependent on $\tau_{CV}^{0}$ and therefore it may not be unique. This uniqueness ambiguity leads to the information paradox such as the unproven theorem paradox.

To eliminate this ambiguity, Deutsch's maximum entropy rule of D-CTCs suggests that one should choose the consistent solution $\tau_{CV}$ with the maximal von Neumann entropy. The existence of such a solution is guaranteed so that the uniqueness ambiguity can be resolved. But there is no physical principle for making such a choice and the maximum entropy rule may not be an essential component of the D-CTCs.

One way to resolve the uniqueness ambiguity of D-CTCs was proposed by Allen in \cite{allen2014treating}, where some noise is incorporated along the path of the CV system by a new quantum channel $N$ so that the consistency condition of the modified circuit is given by
\begin{eqnarray}\label{eq-4}
\tau_{CV}&=&ND_{U}(\rho_{CR}^{in},\tau_{CV})=N(D_{U}(\rho_{CR}^{in},\tau_{CV}))\nonumber\\
&=&pI_{CV}+(1-p)Tr_{CR}(U(\rho_{CR}^{in}\otimes\tau_{CV})U^{\dagger})
\end{eqnarray}

where $N$ is modeled by a depolarization channel with $0<p<1$ and $I_{CV}$ is the maximally mixed state of the CV system.
It was proven that the addition of noise to CV system can solve the uniqueness ambiguity without the maximum entropy rule\cite{allen2014treating}. Using Eq.(\ref{eq_3}), the unique fixed CV system state is $\tau_{CV}^{0}$ independent given by
\begin{equation}\label{eq_5}
\overline{ND_{U}^{\infty}}(\rho_{CR}^{in},\tau_{CV}^{0})=\lim_{n\rightarrow\infty}\frac{1}{n+1}\sum_{k=0}^{n}ND_{U}^{k}(\rho_{CR}^{in},\tau_{CV}^{0})
\end{equation}

\begin{figure}[tbp]
  \centering
  \mbox{
  \subfigure[]
  {\includegraphics[width=2.8cm]{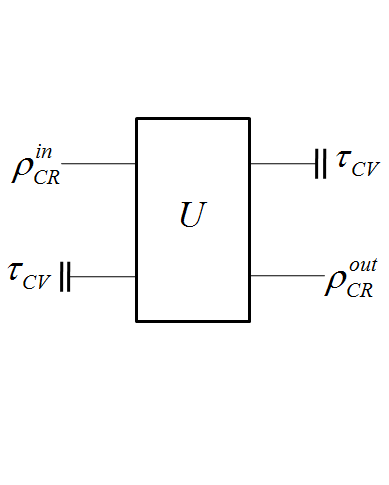}
  \label{fig_01}}
  \hspace{0.3cm}
  \subfigure[]
  {\includegraphics[width=5cm]{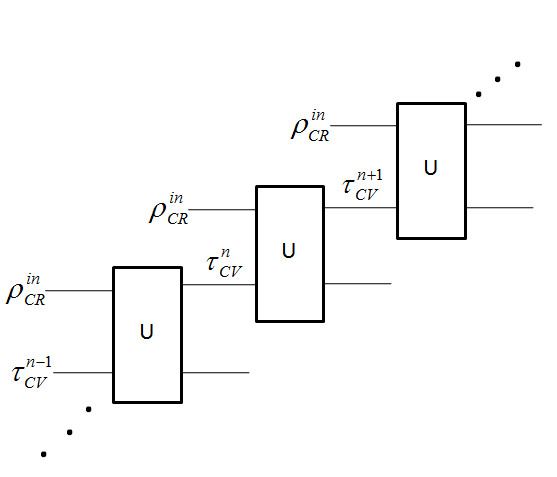}
  \label{fig_02}}
  }
 \caption{The equivalent circuit model of D-CTCs. (a)The D-CTC model with an unitary operation $U$, an initial CR input $\rho_{CR}^{in}$ and a CR output $\rho_{CR}^{out}$. The CV system converges to $\tau_{CV}$ according to the consistency condition of D-CTCs; (b)The equivalent circuit model of the D-CTC system, where the D-CTC circuit is unwrapped into an infinite number of identical ladders of the operation $U$ between $\tau_{CV}^{n}$ and $\rho_{CR}^{in}$\cite{PhysRevA.82.062330}.}\label{fig_equivalent_model}
\end{figure}

An alternative strategy is Ralph and Myers's equivalent circuit model\cite{PhysRevA.82.062330} of D-CTCs, which attempts to derive the maximum entropy rule and resolve the uniqueness ambiguity of D-CTCs. The equivalent circuit model of D-CTCs with an unitary interaction $U$ is constructed by unwrapping the D-CTC circuit and letting the CV system experience an infinite number of unitary interaction $U$ with perfect clones of the CR system along time\cite{PhysRevA.82.062330}\cite{allen2014treating}, as outlined in Fig. \ref{fig_equivalent_model}.

In the equivalent circuit, the CV system state $\tau_{CV}^{n-1}$ evolves with the \emph{n}th ladder of the equivalent circuit as
\begin{equation}\label{eq_6}
  \tau_{CV}^{n}=D_{U}(\rho_{CR}^{in},\tau_{CV}^{n-1})=Tr_{CR}(U(\rho_{CR}^{in}\otimes\tau_{CV}^{n-1})U^{\dagger})
\end{equation}

The output of the equivalent circuit, given in Eq. (\ref{eq_7}), is obtained by executing an infinite ladder of the circuit starting from an initial CV state $\tau_{CV}^{0}$.
\begin{equation}\label{eq_7}
  \tau_{CV}^{out}(\tau_{CV}^{0})=\lim_{n\rightarrow \infty}D_{U}^{n}(\rho_{CR}^{in},\tau_{CV}^{0})
\end{equation}

Similar with the strategy of Allen\cite{allen2014treating}, the equivalent circuit model attempts to solve the uniqueness ambiguity by introducing an arbitrary small decoherence interaction $N$ on the CV system in each iteration of the model. With the decoherence interaction, Eq.(\ref{eq_6}) becomes
\begin{eqnarray}\label{eq_8}
\tau_{CV}^{n}&=&N(D_{U}(\rho_{CR}^{in},\tau_{CV}^{n-1}))\nonumber\\
&=&(1-p)D_{U}(\rho_{CR}^{in},\tau_{CV}^{n-1})+pI_{CV}
\end{eqnarray}
where $0<p\ll 1$ is a small value to control the strength of the decoherence interaction.

Accordingly the output CV system state is now
\begin{eqnarray}\label{eq_9}
 &&\tau_{CV}(U,\rho_{CR}^{in},\tau_{CV}^{0})\nonumber\\
 &=&\lim_{n\rightarrow\infty}[(1-p)^{n}D_{U}^{n}(\rho_{CR}^{in},\tau_{CV}^{0})\nonumber\\
 &+& \sum_{k=0}^{n-1}p(1-p)^{k}D_{U}^{k}(\rho_{CR}^{in},I_{CV})]
\end{eqnarray}

Ralph and Myers\cite{PhysRevA.82.062330} claimed that Detusch's maximum entropy rule can be derived from their model based on the following observations:

\begin{itemize}
  \item Non-unique solutions in the D-CTC model correspond to equivalent circuits that are sensitive to initial conditions of the CV system, $\tau_{CV}^{0}$.

  \item An inclusion of an arbitrarily small decoherence in the equivalent circuit leads to the convergence of the CV system to a unique solution that corresponds to Deutsch's maximum entropy state.

  \item The unique maximum entropy state is obtained from the equivalent circuit as $D_{U}^{\infty}(\rho_{CR}^{in},I_{CV})$ by considering idealized unitaries but starting the iteration from the maximally mixed CV state $I_{CV}$.
\end{itemize}

But the above statements were challenged by Allen in \cite{allen2014treating}. He noticed that the equivalent circuit does not always converge when there is no decoherence interaction. This point was demonstrated by a counterexample in \cite{allen2014treating}. Also the \emph{proof} of \cite{PhysRevA.82.062330}, employed to derive the maximum entropy rule from the equivalent circuit model, is not exact or at least not complete. The key flaw of the \emph{proof} is that it allows every state $\tau_{CV}$ to be a fixed point\cite{allen2014treating}, which is definitely not true.

To this point, we have three pictures of D-CTCs.
\begin{description}
  \item[Deutsch's Picture] The CV system has consistent solutions in the form of Eq.(\ref{eq_3}). When multiple solutions exist, Deutsch's maximum entropy rule is applied to resolve the uniqueness ambiguity and the CV state is the maximum entropy state.
  \item[Allen's picture] The CV system also has consistent solutions given by Eq.(\ref{eq_3}). When multiple solutions exist, incorporating noise in the CV system as in Eq.(\ref{eq-4}) results in a unique solution given by Eq.(\ref{eq_5}) and the maximum entropy rule is not needed.
  \item[Ralph's picture] The CV system can have fixed states as given by Eq.(\ref{eq_7}). The uniqueness ambiguity can be resolved by introducing an arbitrary small decoherence in the CV system (Eq.(\ref{eq_8})). This leads to the unique fixed state Eq.(\ref{eq_9}), which is claimed to coincide with Deutsch's maximum entropy state.
\end{description}

In this paper, we will prove
\begin{itemize}
  \item These three pictures are not equivalent.
  \item Deutsch's maximum entropy rule can not be reproduced by Ralph's or Allen's pictures.
  \item A revised maximum entropy rule can be derived from Ralph's equivalent circuit model.
  \item The revised maximum entropy rule may restore continuous quantum evolutions of D-CTCs.
\end{itemize}

\section{Comparisons of the three pictures of D-CTCs}

Before our discussion on the relationship among the pictuers, we first clarify the following observations:
\begin{itemize}
  \item Eq.(\ref{eq_3})(\ref{eq_5})(\ref{eq_9}) can be guaranteed to converge when $n\rightarrow\infty$, which can be easily proven using Cauchy's criteria of convergence.
  \item  For a given system setup $U,\rho_{CR}^{in},\tau_{CV}^{0},p$, Allen proved that Eq.(\ref{eq_5}) converges to a unique $\tau_{CV}^{0}$ independent state for almost any $0<p<1$ except for the case that $(1-p)^{-1}$ is an eigen value of $D_{U}$\cite{allen2014treating}.  Similarly,  Eq.(\ref{eq_9}) also converges to a $\tau_{CV}^{0}$ indepedent state at $\sum_{k=0}^{\infty}p(1-p)^{k}D_{U}^{k}(\rho_{CR}^{in},I_{CV})$ for any $0<p<1$. We can also verify, for a given $p$ that Eq.(\ref{eq_5}) converges, it will lead to the same unique fixed point as Eq.(\ref{eq_9}).
  \item The converged fixed state of Eq.(\ref{eq_5}) and Eq.(\ref{eq_9}) is $p$ dependent. It fulfills the consistency condition of Eq.(\ref{eq-1}) when $p\rightarrow0$.
\end{itemize}

Among the three pictures of D-CTCs, Ralph's equivalent circuit model possesses the least physical structure ambiguity by apparently unwrapping the CTC path. So we can examine their relationship by taking Ralph's picture as a reference. We can check if the other two pictures can be derived from Ralph's picture with system parameters $U,\rho_{CV}^{in},\tau_{CV}^{0},p$.

\begin{description}
  \item[Case 1 ] When the fixed CV system state of Ralph's picture is unique even without the decoherence interaction, i.e. Eq.(\ref{eq_7}) converges to a unique $\tau_{CV}^{0}$ independent state, it can be verified that Eq.(\ref{eq_3})(\ref{eq_5})(\ref{eq_7}) converge to the same unique consistent CV solution. So we have no uniqueness ambiguity and the three pictures are equivalent. The subspace of unitary operations that may violate this uniqueness assumption is of measure zero\cite{allen2014treating}.
  \item[Case 2] If the equivalent circuit model have multiple fixed CV states when no decoherence is applied, then Eq.(\ref{eq_7}) always converges but it's $\tau_{CV}^{0}$ dependent. Under this assumption, all the three pictures have the same multiple fixed CV states and they suffer from the same uniqueness ambiguity problem. The three pictures are equivalent if we do not consider their different strategies to eliminate the uniqueness ambiguity.
  \item[Case 3] This is the case that the equivalent circuit model does not converge without the decoherence interaction, as pointed out by Allen\cite{allen2014treating}. It's interesting that though Eq.(\ref{eq_7}) does not converge, still Deutsch and Allen's pictures have the same converged fixed CV state as given by Eq.(\ref{eq_3}). Obviously, in this case, Deutsch and Allen's pictures are identical but different with Ralph's picture. We also point out here that even Eq.(\ref{eq_7}) does not converge, introducing a small decoherence to the equivalent circuit model will lead to a convergent state given by Eq.(\ref{eq_9}).

      We illustrate this situation by considering a D-CTC system, in which the CR system is a single qubit and the CV system is a 4-level system. The unitary operation $U_1$ is given by
\begin{eqnarray}\label{eq_10}
 U_1 &=&|00\rangle\langle01|+|10\rangle\langle02|+|02\rangle\langle03|+|03\rangle\langle00|\nonumber\\
 &+&|01\rangle\langle10|+|11\rangle\langle11|+|12\rangle\langle12|+|13\rangle\langle13|
\end{eqnarray}

If the input CR system state is $\rho_{CR}^{in}=|0\rangle\langle0|$, without the decoherence interaction, Ralph's equivalent circuit model may never converge. For example, if the initial CV system state is the maximally mixed state $I_{CV}$, then  $\tau_{CV}=D_{U}^{\infty}(|0\rangle\langle0|,I_{CV})$ will be in a circulation among three states given by $\tau_{CV}^{n-1}=Diag(\frac{1}{2},0,\frac{1}{4},\frac{1}{4})$, $\tau_{CV}^{n}=Diag(\frac{1}{4},0,\frac{1}{4},\frac{1}{2})$, $\tau_{CV}^{n+1}=Diag(\frac{1}{4},0,\frac{1}{2},\frac{1}{4})$. But in Allen's picture, Eq.(\ref{eq_3}),$\frac{1}{n+1}\sum_{k=0}^{n}\tau_{CV}^{k}$ still converges
to the maximum entropy state $\tau_{CV}=Diag(\frac{1}{3},0,\frac{1}{3},\frac{1}{3})$ when $n\rightarrow\infty$. So Allen's picture and Ralph's picture are not equivalent in this case.

\item[Case 4] We now consider how the uniqueness ambiguity can be resolved in the three pictures. We ask the following questions:
    \begin{itemize}
      \item Can the uniqueness ambiguity be resolved successfully in each picture?
      \item Will the three strategies arrive at the same unique consistent CV state so that the maximum entropy rule in Deutsch's picture can be reproduced from Allen and Ralph's pictures?
    \end{itemize}

    Answers to our first question is straightforward. For Deutsch's picture, the existence and uniqueness of the consistent maximum entropy state is ensured\cite{deutsch1991quantum}. For Allen and Ralph's pictures, by introducing noise or decoherence to the CV system, Eq.(\ref{eq_5}) and Eq.(\ref{eq_9}) will converge to the same unique CV state for almost all $0<p<1$ and the converged unique CV state fulfills the consistency condition of Eq.(\ref{eq-1}) if $p\rightarrow0$ as we claimed before. For example, in our example system Eq.(\ref{eq_10}), if noise or a decoherence interaction is applied, then both Allen and Ralph's pictures will converge to $\tau_{CV}=Diag(\frac{1}{3},0,\frac{1}{3},\frac{1}{3})$, which coincides with Deutsch's maximum entropy state.
Therefore the uniqueness ambiguity can be resolved successfully in all the three pictures, Allen and Ralph's pictures are \emph{almost} equivalent to each other except for the case when Allen's picture does not converge for some special values of $p$.

    To answer the second question, we consider another D-CTC system with a 1-qubit CR system and a 4-level CV system. The unitary operation $U_2$ is now
\begin{eqnarray}\label{eq_D_CTC6}
U_2 &=&|10\rangle\langle00|+|00\rangle\langle01|+\frac{1}{\sqrt{2}}(|02\rangle+|13\rangle)\langle02|\nonumber\\
&+&\frac{1}{\sqrt{2}}(|03\rangle+|12\rangle)\langle03|+|11\rangle\langle10|+|01\rangle\langle11|\\
&+&\frac{1}{\sqrt{2}}(|13\rangle-|02\rangle)\langle12|+\frac{1}{\sqrt{2}}(|03\rangle-|12\rangle)\langle13|\nonumber
\end{eqnarray}

 When the input CR state is $\rho_{CR}^{in}=|0\rangle\langle0|$, there exist at least two consistent CV states, which are $\tau_{AV}^{out1}=\frac{1}{2}|0\rangle\langle0|+\frac{1}{4}|2\rangle\langle2|+\frac{1}{4}|3\rangle\langle3|$ with a von Neumann entropy of $1.5$ bits and $\tau_{CV}^{out2}=\frac{1}{3}|0\rangle\langle0|+\frac{1}{3}|2\rangle\langle2|+\frac{1}{3}|3\rangle\langle3|$ with a von Neumann entropy of $1.585$ bits. $\tau_{CV}^{out2}$ is the maximum entropy state for the fixed $\rho_{CR}^{in}=|0\rangle\langle0|$ and $U_2$.

We can easily verify that
\begin{itemize}
  \item Taking the initial CV system state as the maximal mixed state $\tau_{CV}^{01}=\frac{1}{4}(|0\rangle\langle0|+|1\rangle\langle1|+|2\rangle\langle2|+|3\rangle\langle3|)$, the equivalent circuit model will converge to the non-maximum entropy fixed state $\tau_{CV}^{out1}$, no matter if a small decoherence is at present or not.
  \item Interestingly, the equivalent circuit model starting from the initial CV state $\tau_{CV}^{02}=\frac{1}{3}|0\rangle\langle0|+\frac{1}{3}|2\rangle\langle2|+\frac{1}{3}|3\rangle\langle3|$ will converge to the maximum entropy state $\tau_{CV}^{out2}$ without the decoherence. But it still converges to the non-maximum entropy state $\tau_{CV}^{out1}$ when a small decoherence exists in each iteration of the equivalent circuit model.
\end{itemize}

For the case of $\rho_{CR}^{in}=|0\rangle\langle0|$, we can draw the following conclusions
\begin{itemize}
  \item The maximum entropy state $\tau_{CV}^{out2}$ is \emph{unstable} with respect to the decoherence interaction of the equivalent circuit model and it can only be reached by the equivalent circuit model without any decoherence.
  \item The non-maximum entropy state $\tau_{CV}^{out1}$ is \emph{stable} and the maximum entropy state is prone to \emph{collapse} to $\tau_{CV}^{out1}$ when the decoherence interaction is introduced in the equivalent circuit model.
  \item The maximum entropy state $\tau_{CV}^{out2}$ is not generated from the maximally mixed CV system state $\tau_{CV}^{0}=I_{CV}$.
  \item The equivalent circuit model with the decoherence interaction does not always reproduce the maximum entropy rule of D-CTCs in this example.
\end{itemize}

\begin{figure}[tbp]
  \centering
  \mbox{
  \subfigure[]
  {\includegraphics[width=4.2cm]{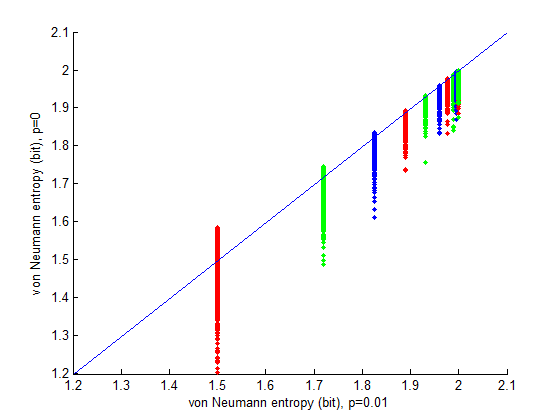}
  \label{fig_21}}

  \subfigure[]
  {\includegraphics[width=4.2cm]{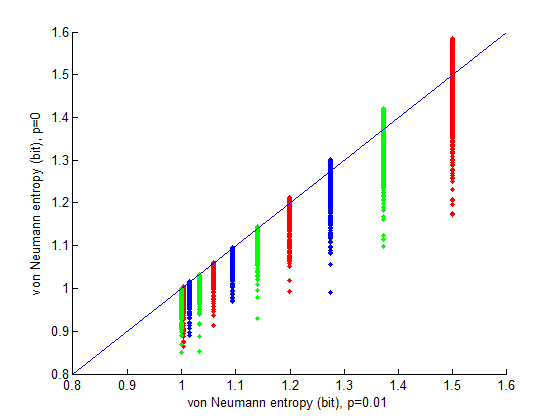}
  \label{fig_22}}
  }

 \caption{Simulation results of the equivalent circuit model of a D-CTC system with the unitary operation $U_2$. The decoherence factor $p$ is taken as $p=0$ for the case of no decoherence and $p=0.01$ for a small decoherence case. For each given input state of $\rho_{CR}^{in}$, 1000 random initial states $\tau_{CV}^{0}$ are simulated, which are fed to the equivalent circuit model for 10000 iterations to reach converged $\tau_{CV}$ states. For each data point, its x axis and y axis give the von Neumann entropy of the converged CV state for $p=0$ and $p=0.01$ respectively. The data points above the diagonal blue lines indicate the cases that the maximum entropy states can not be reproduced by the equivalent circuit model with a small decoherence. (a)The von Neumann entropies of the converged CV system outputs using the equivalent circuit model with mixed CR states $\rho_{CR}^{in}= \frac{1}{1+s}(|0\rangle\langle 0|+s|1\rangle\langle 1|)$, $s=0,0.1,0.2,...1$ (from left to right) and random initial CV states.  (b)The von Neumann entropies of the converged CV system outputs of the equivalent circuit model with pure CR input states $\rho_{CR}^{in}= \frac{1}{1+s^2}(|0\rangle+s|1\rangle)(\langle0|+s\langle1|)$, $s=0,0.1,0.2,...1$ (from right to left) and random initial CV states.}\label{fig_2}
\end{figure}

In order to further clarify that the equivalent circuit model of our example system does not only fail to reproduce the maximum entropy rule for the special case $\rho_{CR}^{in}=|0\rangle\langle0|$, we carried out a simulation study on the equivalent circuit model with different CR input states $\rho_{CR}^{in}$ and random initial CV system states $\tau_{CV}^{0}$. From the simulation results in Fig. \ref{fig_2}, we see that

 \begin{itemize}
   \item For all the different pure and mixed input CR system states in our simulation, there exist multiple fixed states of the CV system. Introducing a small decoherence interaction does lead to a unique fixed state of $\tau_{CV}$ for each fixed input CR system state $\rho_{CR}^{in}$.
   \item For each given $\rho_{CR}^{in}$, there exist some initial states $\tau_{CV}^{0}$ (shown by the data points above the diagonal blue lines in Fig. \ref{fig_2}) so that the equivalent circuit model with the decoherence will not converge to the maximum entropy CV state, since the correspondent fixed state without the decoherence interaction has a larger von Neumann entropy.
 \end{itemize}

Our counterexample can also help to confirm Allen's suspicion\cite{allen2014treating} on the incompleteness of Ralph's work in \cite{PhysRevA.82.062330}, where the convergence of the equivalent circuit model (Eq. (\ref{eq_7})) is represented by a Kraus decomposition as
\begin{eqnarray}\label{eq_12}
  \tau_{CV}^{out}(\tau_{CV}^{0})&=&\lim_{n\rightarrow \infty}D_{U}^{n}(\rho_{CR}^{in},\tau_{CV}^{0}) \nonumber\\
   &=& \sum_{j}E_{j}\tau_{CV}^{0}E_{j}^{\dagger}\\
   1 &=& \sum_{j}E_{j}^{\dagger}E_{j}
\end{eqnarray}
Then as a fixed point of the equivalent circuit model, $\tau_{CV}^{out}(\tau_{CV}^{0})=\sum_{j}E_{j}\tau_{CV}^{0}E_{j}^{\dagger}$ fulfills
\begin{eqnarray}\label{eq_13}
 \sum_{j}E_{j}\tau_{CV}^{0}E_{j}^{\dagger}&=&\sum_{j,k}E_{j}E_{k}\tau_{CV}^{0}E_{k}^{\dagger}E_{j}^{\dagger}\\
 \sum_{j,k}E_{j}E_{k}^{\dagger}E_{k}\tau_{CV}^{0}E_{j}^{\dagger}&=&\sum_{j,k}E_{j}E_{k}\tau_{CV}^{0}E_{k}^{\dagger}E_{j}^{\dagger}
\end{eqnarray}

Ralph then claimed that Eq.(\ref{eq_13}) implies
\begin{equation}\label{eq_14}
[E_{j}\tau_{CV}^{0},E_{j}^{\dagger}]=0
\end{equation}
This is the key component for the deduction of the maximum entropy role from the equivalent circuit model in \cite{PhysRevA.82.062330}.

Allen questioned the validness of Eq.(\ref{eq_14}) but he did not give an explicit counterexample of it\cite{allen2014treating}. With our second example, it can be verified that for $\rho_{CR}^{in}=|0\rangle\langle0|$, the equivalent circuit model converges and the correspondent Kraus operators $\{E_{j}\},j=1,2,3,4$ are given by
\begin{equation}\label{eq_15}
 E_{1}=\left[
  \begin{array}{cccc}
  1 & 0 & 0 & 0 \\
  0 & 0 & 0 & 0 \\
  0 & 0 & 0 & 0 \\
  0 & 0 & 0 & 0
  \end{array}\right]
 E_{2}=\left[
  \begin{array}{cccc}
  0 & 1 & 0 & 0 \\
  0 & 0 & 0 & 0 \\
  0 & 0 & 0 & 0 \\
  0 & 0 & 0 & 0
  \end{array}\right]
\end{equation}

\begin{equation}\label{eq_16}
 E_{3}=\left[
  \begin{array}{cccc}
  0 & 0 & 0 & 0 \\
  0 & 0 & 0 & 0 \\
  0 & 0 & \frac{1}{\sqrt{2}} & 0 \\
  0 & 0 & 0 & \frac{1}{\sqrt{2}}
  \end{array}\right]
 E_{4}=\left[
  \begin{array}{cccc}
  0 & 0 & 0 & 0 \\
  0 & 0 & 0 & 0 \\
  0 & 0 & 0 & \frac{1}{\sqrt{2}} \\
  0 & 0 & \frac{1}{\sqrt{2}} & 0
  \end{array}\right]
\end{equation}
Obviously, Eq.(\ref{eq_14}) does not hold for $\{E_{j}\},j=1,2,3,4$ and Ralph's derivation of Deutsch's maximum entropy role from the equivalent circuit model is wrong.
\end{description}

\section{Revised maximum entropy rule and discontinuous quantum evolutions in D-CTCs}
Though our explicitly constructed examples tell us that the equivalent circuit model is not identical with Deutsch's D-CTC model, this does not necessarily mean that the equivalent circuit model is not the proper model of D-CTCs and therefore should be abandoned. Instead, we even incline to assign this inconsistency between them as a hint that the original D-CTC model is not exact or realistic because
\begin{itemize}
  \item In our second example, the maximum entropy state $\tau_{CV}^{out2}$ can only be achieved by the equivalent circuit model with no decoherence interaction. But the decoherence is inevitable and therefore the maximum entropy state is unstable. A realistic system with a decoherence interaction will converge to a \emph{stable maximum entropy state} instead of an \emph{unstable maximum entropy state}.

  \item The equivalent circuit model has succeeded in helping us to clarify the properties of D-CTCs dealing with proper and improper mixtures such as the \emph{linear trap} argument of \cite{bennett2009can} for the non-orthogonal states discrimination\cite{PhysRevA.82.062330}. This point was also partially supported by simulated experiments on D-CTCs\cite{ringbauer2014experimental}.

  \item It's observed in \cite{pati4221purification} that it's impossible to purify a mixed CV state in D-CTCs.  In Ralph's picture, this can be intuitively explained as that the CV system can be entangled with an infinite number of copies of the CR system by iteratively applying the unitary interaction $U$ in the equivalent circuit (see Fig. \ref{fig_equivalent_model}). Therefore the mixed CV state can only be purified by an infinite sized ancillary system and the universal 'Chruch of the larger Hilbert space' does not exist.
\end{itemize}

So we speculate that as a heuristic rule to describe the behaviour of D-CTCs, the maximum entropy rule of D-CTCs can be modified as: \textbf{When multiple fixed states of the CV system exist, the D-CTCs will choose the \emph{stable} maximum entropy state determined by Ralph's equivalent circuit model with a decoherence interaction. But the final fixed CV state might be decoherence interaction dependent}.

Now we use our revised maximum entropy rule to resolve discontinuous quantum evolutions of D-CTCs, which is regarded as one essential property of D-CTCs. In \cite{dejonghe2010discontinuous} an explicit example was constructed to demonstrate that D-CTCs allow both ephemerally and physically discontinuous gates. Here we show that both the discontinuities on the CV and CR systems may be eliminated by our revised maximum entropy rule.

We examine the same example of \cite{dejonghe2010discontinuous}, where the CR system includes 2 qubits as $\rho_{CV}^{in}=\rho_{\alpha}\otimes\rho_{\beta}$ and the CV system is a single qubit. The unitary operator is given by
\begin{eqnarray}
  U&=&|000\rangle\langle100|+|001\rangle\langle001|+|010\rangle\langle011|+|011\rangle\langle010|\nonumber\\
   &+&|100\rangle\langle000|+|101\rangle\langle110|+|110\rangle\langle101|+|111\rangle\langle111|\nonumber
\end{eqnarray}\label{eq_18}
For three CR system states $\rho_{CV}^{A},\rho_{CV}^{B},\rho_{CV}^{C}$, where $(\rho_{\alpha}^{A})_{11}=1,(\rho_{\beta}^{A})_{11}=1-\epsilon_{\beta}$, $\rho_{\alpha}^{B}=\rho_{\beta}^{B}=|0\rangle\langle0|$ and $(\rho_{\alpha}^{C})_{11}=1-\epsilon_\alpha,(\rho_{\beta}^{C})_{11}=1$ with $0<\epsilon_{\alpha},\epsilon_{\beta}\ll1$ , the correspondent consistent CV system states are
$\tau_{CV}^{A}=\{\tau_{CV}|(\tau_{CV})_{12}=0\}$, $\tau_{CV}^{B}=\frac{1}{2}(|0\rangle\langle0|+|1\rangle\langle1|)$ and $\tau_{CV}^{C}=|0\rangle\langle0|$. Then as $\epsilon_{\alpha},\epsilon_{\beta}\rightarrow0$, the three initial CR states are arbitrary close but the correspondent CV system states are discontinuous. When Deutsch's maximum entropy rule is applied, the discontinuity between $\tau_{CV}^{A}$ and $\tau_{CV}^{B}$ can be eliminated as $\tau_{CV}^{A}=\tau_{CV}^{B}=\frac{1}{2}(|0\rangle\langle0|+|1\rangle\langle1|)$, but there remains a discontinuity with $\tau_{CV}^{C}$ at $\epsilon_{\alpha}=0$.

\begin{figure}[tbp]
  \centering
  \mbox{
  \subfigure[]
  {\includegraphics[width=4.2cm]{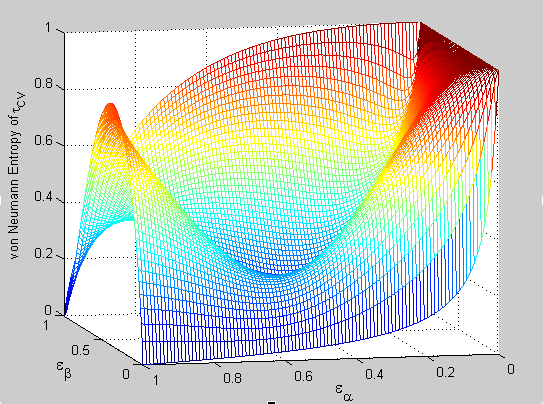}
  \label{fig_33}}

  \subfigure[]
  {\includegraphics[width=4.2cm]{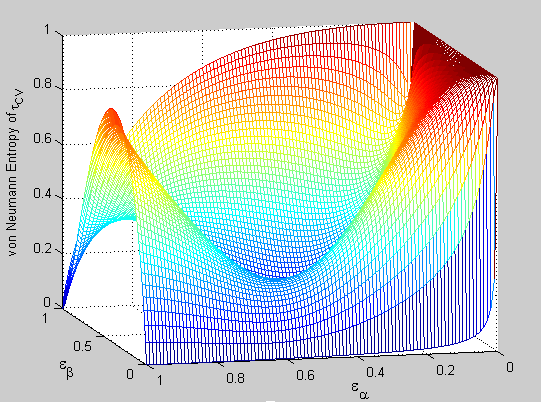}
  \label{fig_34}}
  }

  \mbox{
  \subfigure[]
  {\includegraphics[width=4.2cm]{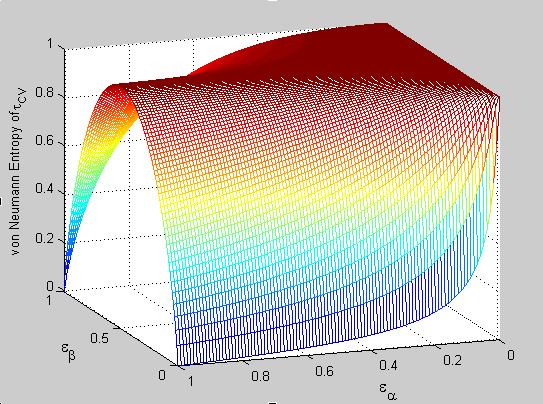}
  \label{fig_31}}

  \subfigure[]
  {\includegraphics[width=4.2cm]{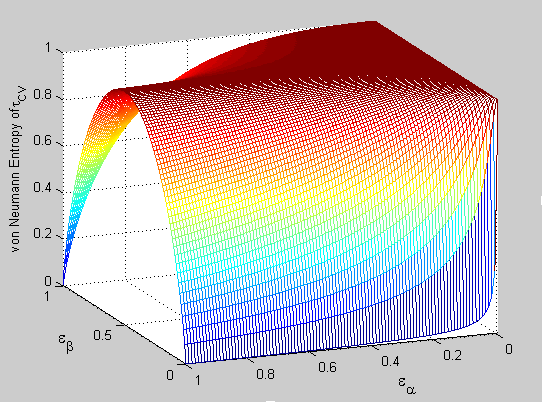}
  \label{fig_32}}
  }

 \caption{von Neumann entropy of the unique consistent CV system state determined by the revised maximum entropy rule generated from Ralph's equivalent circuit model of D-CTCs. The smooth surfaces of von Neumann entropy demonstrate that continuous quantum evolutions are restored in D-CTCs. (a)(b)Initial CR system states are pure states and $p=0.1,0.001$; (c)(d)Initial CR system states are mixed states and $p=0.1,0.001$.}\label{fig_3}
\end{figure}

By our revised maximum entropy rule, the consistent CV state should be the stable maximum entropy state determined by the equivalent circuit model with an decoherence interaction. We show this will result in continuous quantum evolutions of both the CR and CV systems.
We choose the initial CR system states at mixed states $\rho_{\alpha}=Diag(1-\epsilon_{\alpha},\epsilon_{\alpha}),\rho_{\beta}=Diag(1-\epsilon_{\beta},\epsilon_{\beta})$ or pure states  $\rho_{\alpha}=(\sqrt{1-\epsilon_{\alpha}}|0\rangle+\sqrt{\epsilon_{\alpha}}|1\rangle)(\sqrt{1-\epsilon_{\alpha}}\langle0|+\sqrt{\epsilon_{\alpha}}\langle1|),
\rho_{\beta}=(\sqrt{1-\epsilon_{\beta}}|0\rangle+\sqrt{\epsilon_{\beta}}|1\rangle)(\sqrt{1-\epsilon_{\beta}}\langle0|+\sqrt{\epsilon_{\beta}}\langle1|)$. Obviously a subset of $\rho_{CV}^{A},\rho_{CV}^{B},\rho_{CV}^{C}$ are special cases of these configurations. The von Neumann entropies of the consistent CV system states for $0\leq\epsilon_{\alpha},\epsilon_{\beta}\leq1$ are reported in Fig. \ref{fig_3}. We can see that the abrupt discontinuity between $\tau_{CV}^{B}$($\epsilon_{\alpha}=\epsilon_{\beta}=0$) and $\tau_{CV}^{C}$($\epsilon_{\alpha}\neq 0,\epsilon_{\beta}=0$) with Deutsch's maximum entropy rule is now replaced by continuous evolutions with our revised maximum entropy rule. Furthermore, a direct computation of Eq.(\ref{eq_9}) on $\rho_{CV}^{in}=\rho_{\alpha}\otimes\rho_{\beta}$ with
\begin{equation}\label{eq_19}
 \rho_{\alpha}=\left[
  \begin{array}{cc}
   1-\epsilon_{\alpha} & \delta_{\alpha} \\
  \delta_{\alpha}^{*} & \epsilon_{\alpha}
  \end{array}\right],  \vspace{0.5cm}
 \rho_{\beta}=\left[
  \begin{array}{cc}
   1-\epsilon_{\beta} & \delta_{\beta} \\
  \delta_{\beta}^{*} & \epsilon_{\beta}
  \end{array}\right]
\end{equation}
 results in continuous evolutions of both the CR and CV systems with respect to $\epsilon_{\alpha},\delta_{\alpha},\epsilon_{\beta},\delta_{\beta}$ and $p>0$. Thought we do not give a complete proof of this point, the above example strongly suggests continuous evolutions of D-CTCs may be restored by Ralph's equivalent circuit model if an arbitrarily small decoherence interaction is involved.

\section{Conclusive remarks}
In this paper we examined the relationship between Deutsch's maximum entropy rule and Ralph's equivalent circuit model of D-CTCs. We first proved the convergency of the equivalent circuit model when an arbitrary small decoherence interaction is introduced in the model. Then by constructing explicit examples, we demonstrated that there exist D-CTC systems whose equivalent circuit models do not exactly reproduce Deutsch's maximum entropy rule and therefore confirmed the suspicion of \cite{allen2014treating} on the incompleteness of the proof of \cite{PhysRevA.82.062330}. Based on these observations, we claim that Ralph's equivalent circuit model will lead to a revised maximum entropy rule. We also suggest that the revised rule might restore continuous quantum evolutions with D-CTCs.
\acknowledgments{
The autherors would like to thank Norbert Schuch for helpful discussion and providing counterexamples to stimulate this work on physics.stackexchange.com. This work was supported in part (H. C.) by Chinese National Science Foundation under Grant No. 61170321.
}

\bibliographystyle{apsrev4-1}
\bibliography{CTC}

%
%
%
%
%
%
%
%
%
%
%
%
%
%
%
%

\end{document}